\documentclass[a4paper,11pt]{article}
\pdfoutput=1 

\usepackage{jinstpub} 
\usepackage{multirow}
\usepackage{graphicx}
\usepackage{lineno}
\usepackage{ulem}
\usepackage{braket}
\usepackage{indentfirst}
\usepackage{bm}

\title{Scintillation light increase of carbontetrafluoride gas at low temperature}

\author[a,1]{Keita Mizukoshi\note{Corresponding author.}}
\author[a]{Takeshi Maeda}
\author[a,2]{Yuuki Nakano \note{Present address: Kamioka Observatory, Institute for Cosmic Ray Research, The University of Tokyo, Gifu 506-1205, Japan}}
\author[a]{Satoshi Higashino}
\author[a]{Kentaro Miuchi}

\affiliation[a]{Graduate School of Science, Kobe University, 1-1, Rokkodai, Nada, Kobe 657-0013, Japan}
\emailAdd{mzks@stu.kobe-u.ac.jp}

\begin{document}

\abstract{Scintillation detector is widely used for the particle detection in the field of particle physics. 
Particle detectors containing fluorine-19~($^{19}\mathrm{F}$) are known to have advantages for Weakly Interacting Massive Particles~(WIMPs) dark matter search, especially for spin-dependent interactions with WIMPs due to its spin structure. 
In this study, the scintillation properties of carbontetrafluoride~($\mathrm{CF_{4}}$) gas at low temperature was evaluated because its temperature dependence of light yield has not been measured.
We evaluated the light yield by cooling the gas from room temperature~(300~K) to 263~K. 
As a result, the light yield of $\mathrm{CF_{4}}$ was found to increase by $(41.0\pm4.0_{\rm stat.}\pm6.6_{\rm syst.})\%$ and the energy resolution was also found to improve at low temperature.}

\maketitle
\flushbottom


\section{Introduction} \label{s1_intro}

Scintillation detectors have been one of the  
most widely-used instruments 
for particle detection. 
Since large detectors can be built at a relatively low cost, they have been used in a wide range of applications for rare-event-search  experiments, such as dark matter searches~\cite{xenon, xmass, DEAP:2019yzn}, neutrinoless double beta decay searches~\cite{kamland, CANDLES:2020iya}, and studies of neutrino interactions~\cite{borexino, juno}.
A high light yield is 
one of the most important properties of these scintillation detectors to realize a low threshold and good energy resolution.
The light yield of the scintillators is known to be temperature dependent in addition to their inherent light production.
Three main mechanisms,~(A) the population of occupied excited levels of electrons,~(B) temperature quenching, and~(C) capture of excited electrons in traps, are known to cause the temperature dependence.~(A) is known to provide a light yield maximum at some temperatures.~(B) gives a monotonic light yield increase with the temperature decrease, while
~(C) gives monotonic light yield decrease with the temperature decrease.
As a result, there is no comprehensive knowledge of how to anticipate the precise temperature dependency of each scintillator, and each one must be evaluated separately.

Particle detectors containing fluorine-19~($^{19}\mathrm{F}$)
are known to have advantages for
Weakly Interacting Massive Particles~(WIMPs) dark matter search~\cite{spin} as well as for (solar and supernova) neutrino detection~\cite{Barabanov:1994rln, munu}.
In particular, the spin structure of  the $^{19}\mathrm{F}$ nucleus enhances the signal from spin-dependent~(SD) interactions with dark matter.
The PICO-60 experiment reported the world-best limit of $3.2 \times 10^{-41}~\mathrm{cm}^{2}$ for $25~\mathrm{GeV}/c^{2}$ WIMPs
with a superheated droplet detector~(SDD)
filled with ~$\mathrm{C_{3}F_{8}}$~\cite{pico}.
While leading experiments can potentially discover WIMPs at any time, it should be noted that SDD is a threshold-type detector, which means the energy spectrum cannot be obtained from one measurement. Since the energy spectrum is thought to be one of the most important pieces of information for the discovery and further studies of WIMPs, it is  well-justified to develop spectroscopy detectors that contain 
 $^{19}\mathrm{F}$.
This type of ``$^{19}\mathrm{F}$ spectroscopy detector'' 
can be realized in the forms of 
 gas time projection chambers~(TPC)~\cite{Tanimori:2003xs, Battat:2014van}, crystal scintillators~\cite{Shimizu:2005kf, Ogawa:2004fy}, and bolometers~\cite{Miuchi:2002zp, Takeda:2003km}.
Among these possibilities, we focused on the carbontetrafluoride ($\rm CF_4$) gaseous scintillator mainly because of the purification feasibility. 
Large mass detectors can be realized once 
$\rm CF_4$ is confirmed to be used as a practical scintillator in the liquid-state in terms of the intrinsic light yield and the self-absorption, which is a future important work of ours.
The properties of gas $\rm CF_4$ scintillation light, such as emission spectra~\cite{VANSPRANG197851, Pansky:1994zh, BROGGINI1995543, cf4prop, cf4prop2, cf4prop3, secondary}, and decay time~\cite{cf4decay, decaytime}, have been investigated for several decades.  
The temperature dependence of its scintillation light yield has not been investigated yet, since TPC detectors with $\mathrm{CF_{4}}$ gas are 
ordinarily operated at the room temperature. 
For this purpose, 
we evaluated the scintillation properties of $\mathrm{CF_{4}}$ gas at a low temperature and the results are reported in this paper.

This paper consists of four sections, including the Introduction. In Sec.~2, we describe the experimental setup to evaluate the temperature dependence of scintillation light from $\rm CF_4$ gas and briefly explain the analysis method. In Sec.~3, we present the experimental results and discussion. The study is concluded in Sec.~4.
\section{Measurements} \label{s2_setup}

\subsection{Setup}
\begin{figure}[t] 
\centering
 \begin{minipage}{0.48\columnwidth}
 \centering
 \includegraphics[width=\columnwidth]{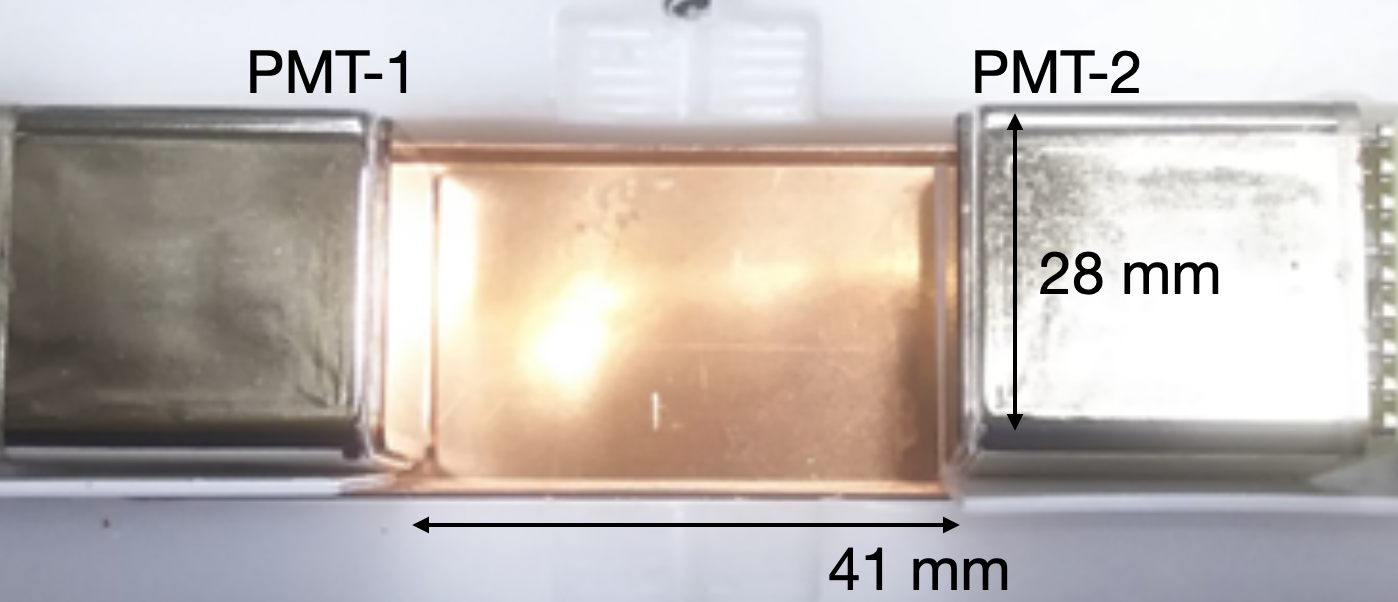}
 \end{minipage}
 \begin{minipage}{0.48\columnwidth}
 \centering
 \includegraphics[width=\columnwidth]{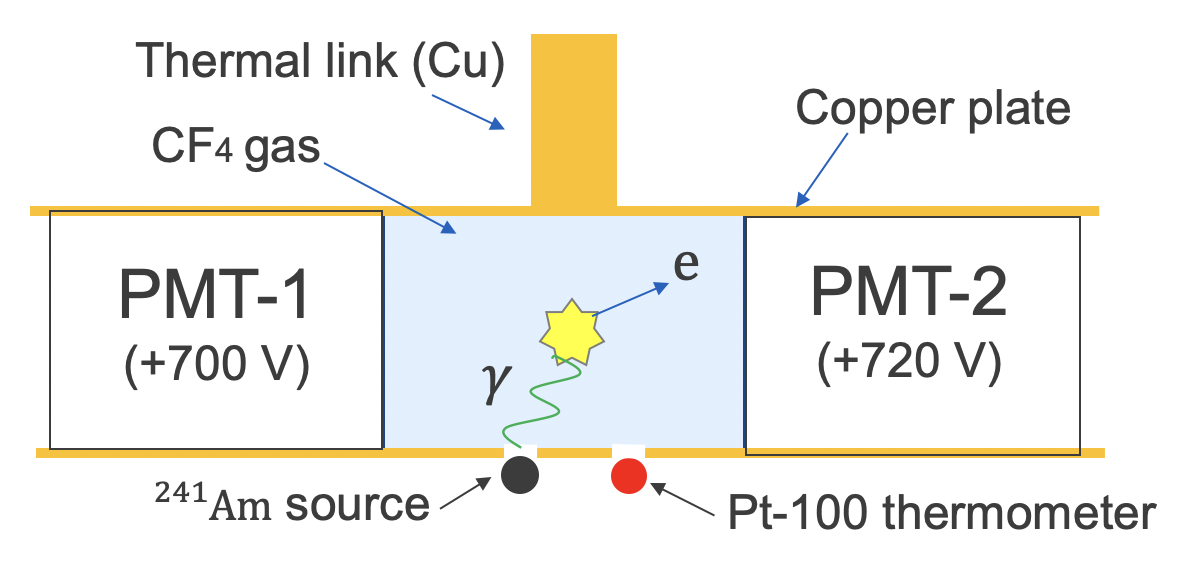}
 \end{minipage}
  \caption{(Left) The top view of the detector. (Right) Conceptional image of the detector setup. The source of $\mathrm{^{241}Am}$ and the thermometer are mounted on the external side of the copper plate. The $\mathrm{^{241}Am}$ source is located at a slightly off-centered position. The distance to one PMT is 14.5~mm while the other is 26.5~mm. It was confirmed that there was no effect of this displacement for the light yield measurement.}
 \label{setup1}
\end{figure}

\begin{figure}[t]
  \begin{center}
    \includegraphics[width=12.0cm]{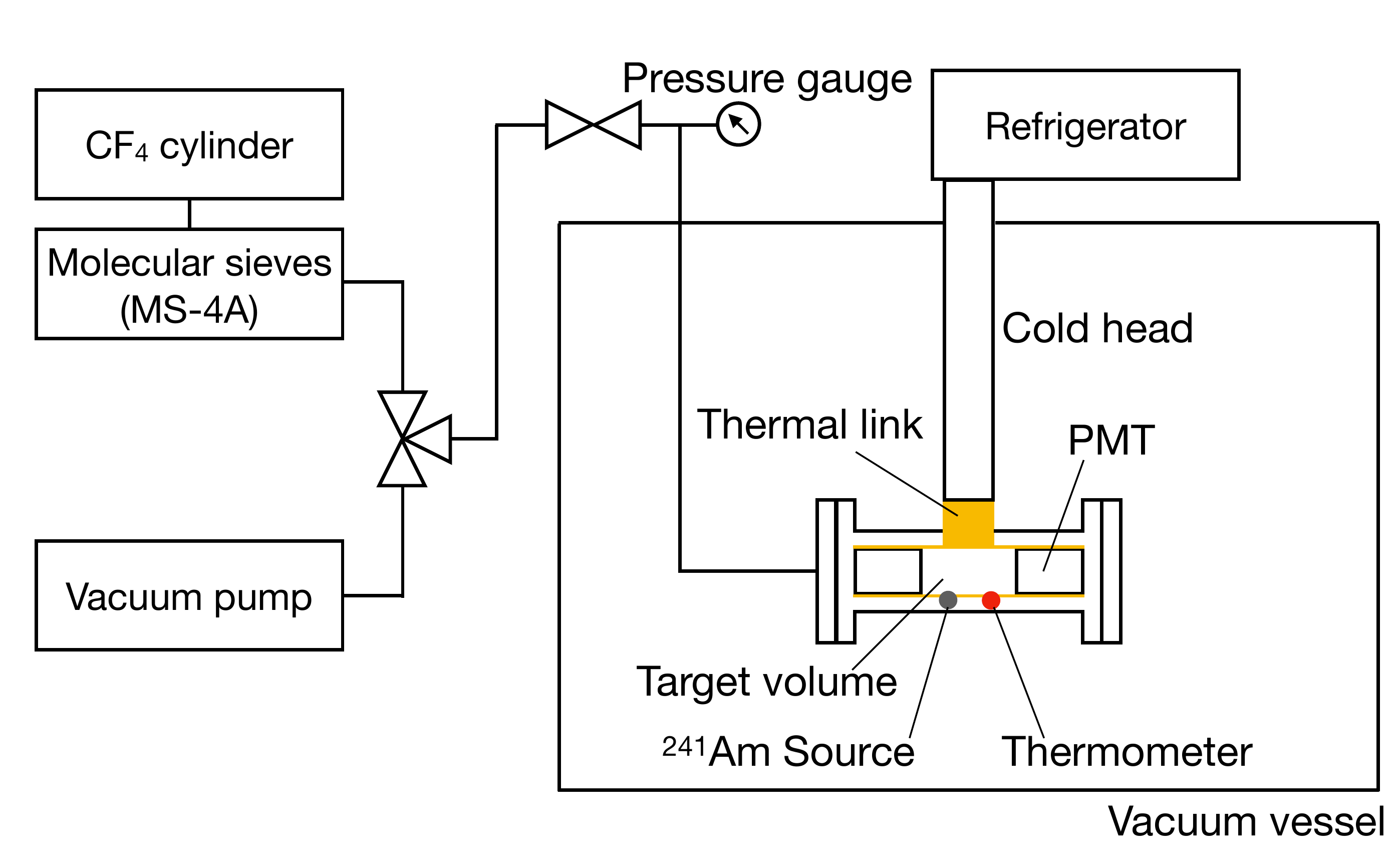}
    \caption{The schematic view of the whole experimental equipment. 
    The detector~(shown in Figure~\ref{setup1}) was set in a vacuum vessel. The detector was cooled down by the refrigerator through the thermal link. The target volume is connected to the vacuum pump and the $\mathrm{CF_{4}}$ gas cylinder.}
    \label{fig:vessel}
  \end{center}
\end{figure}

The detector used for this work was 
constructed at Kobe University. 
A photograph and a conceptional image  are 
shown in Figure~\ref{setup1}.
The target volume of $28\times28\times41~\rm{{mm}^3}$ is viewed by two photomultiplier tubes~(PMTs; PMT-1 and PMT-2) from two sides. 
The other four faces are surrounded by copper plates of $1$~mm thickness.
To preserve the thermal contact between the target gas and the copper plates as high as possible, no extra material to improve reflection is put inside the copper.
An $\mathrm{^{241}Am}$ source~($37$~Bq) is placed on the external side of one of the copper plates, which has a small hole to let $\gamma$-rays~($59.5$~keV) enter into the target volume. The temperature and pressure of the gas were measured by a Pt--100 thermometer~(P0K1.232.6W.B.007, IST INNOVATIVE SENSOR TECHNOLOGY) and a pressure gauge~(MPS--R33RC--NGAT, Myotoku), respectively. 
PMT R8520-406 produced by Hamamatsu K.K. was selected because it was demonstrated to work at a cold temperature of $163~\mathrm{K}$~\cite{datasheet}.
The PMT-1 and PMT-2 were biased at $+700$~V and $+720$~V, respectively, to function at comparable gains. The waveforms of PMTs' signals were inverted with a signal transformer and 
recorded by a DRS4 Evaluation board~\cite{drs4}.
The DRS4 Evaluation board recorded the waveforms 
with a 14-bit precision at  
1 GHz sampling for a range from $-100$~mV to $+900$~mV. The board has a 50~$\Omega$ termination.
The trigger was issued by a coincidence of the two PMTs with a threshold of $+95$~mV.
Figure \ref{fig:vessel} shows the whole experimental equipment.
The detector was set in a stainless cylindrical container with an inner diameter of 60~mm. 
The stainless container had ICF~144 flanges on both ends and was filled with CF$_4$ gas for the measurement.
It was set in a vacuum vessel for thermal insulation.
The target volume was filled with  $\mathrm{CF_{4}}$ gas~(purity grade~5N, $99.999\%$) filtered with 4A-type molecular sieves~(MS-4A).
A refrigerator, PDC08 supplied by Ulvac Cryogenics Inc.~($14$~W capacity), was connected to the copper plate via a copper thermal link.

\subsection{Measurement}
\begin{figure}[t]
  \begin{center}
    \includegraphics[width=\hsize]{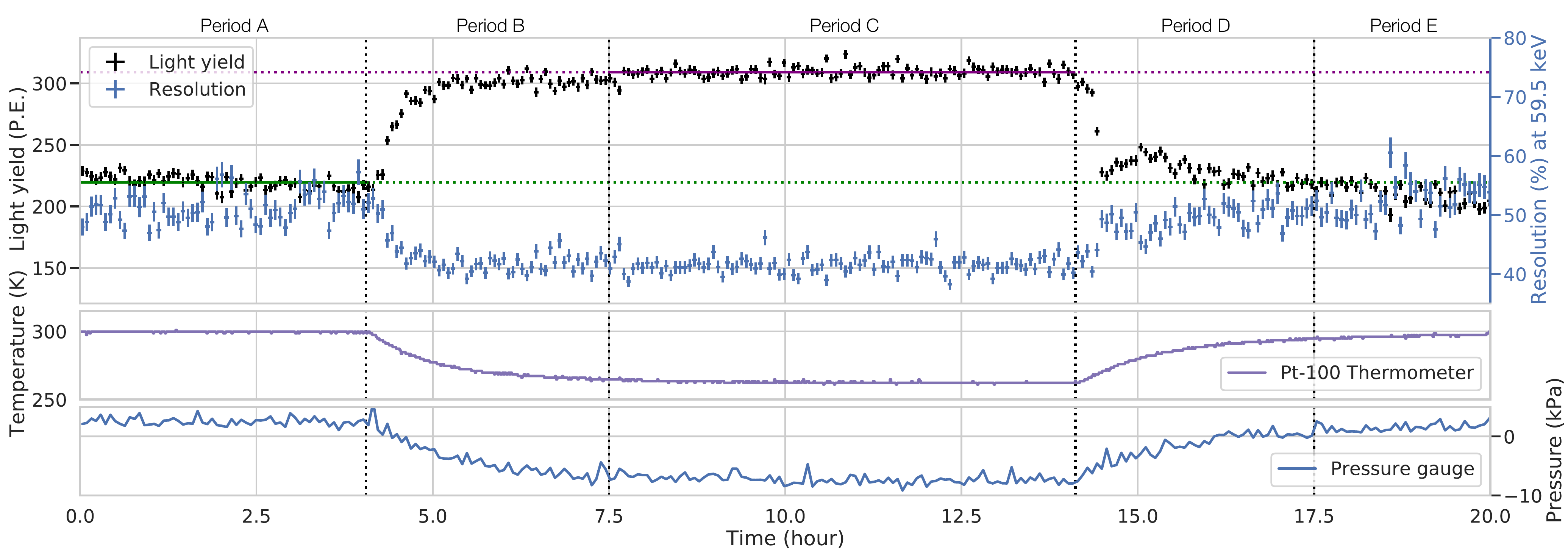}
    \caption{The number of photoelectrons and temperature during the experiment. 
    The top figure shows transitions of the number of photoelectrons. 
    The middle and bottom figures show the temperature and pressure of the CF$_4$ gas, respectively.
    The green and purple lines are mean of the light yield during the stable conditions.
    }
    \label{fig:pesum}
  \end{center}
\end{figure}

The measurement was carried out as one heat cycle starting at the room temperature, cooling down to a low temperature, and then warming-up to the room temperature.
The scintillation signal  was continuously measured throughout the heat cycle at various temperatures. 
First, the target volume was 
evacuated down to $1$~Pa and then filled with $\mathrm{CF_{4}}$ gas 
at room temperature. Soon after the gas pressure reached $1.0 \times 10^5$~Pa, the detector valve was closed and the bias voltages were applied to the PMTs.
The data acquisition was started at this timing, which is defined as the origin of the experimental time~($t_{0}$).
\begin{table}[t]
\caption{Definition of the experimental periods. The term~(hours) is the elapsed time since the start of the measurement~($t_{0}$ in the main text).}
\label{tab:period}
\centering
\begin{tabular}{|ccc|}
\hline 
Period & Hours since t0 & Description                           \\
\hline
A      & $\phantom{00.0}0$ -- $\phantom{0}4.05$ & Room temperature (stable)\\
B      & $\phantom{0}4.05$ -- $\phantom{0}7.50$ & Cooling                               \\
C      & $\phantom{0}7.50$ -- $14.12$ & Cold (stable)                          \\
D      & $14.12$ -- $17.50$ & Warming  \\
E      & $17.50$ -- $20.00$ & Room temperature (stable)\\
\hline
\end{tabular}
\end{table}
The detector temperature and pressure 
during the measurement are shown in the middle and bottom plots of Figure \ref{fig:pesum}.
Five ``Periods'' are defined during the measurement to describe the different conditions as summarized in Table~\ref{tab:period}.
The data was collected at room temperature for four hours as a reference and to evaluate the stability of the measurement~(Period A). We then switched on the refrigerator to start cooling the gas, and data was collected throughout this time~(Period B).
Data was collected at a cold temperature of $263$~K when the instruments attained a steady temperature and gas pressure at which period the thermal uniformity of the gas in the target volume can be enough assumed~(Period C).
After about seven hours of data taking at the cold temperature, the refrigerator was turned off. A heat leak increased the temperature of the gas~(Period D).
At the end, the detector returned to the room temperature and the data was taken for another two hours~(Period E).

\begin{figure}[t]
  \begin{center}
    \includegraphics[width=11.0cm]{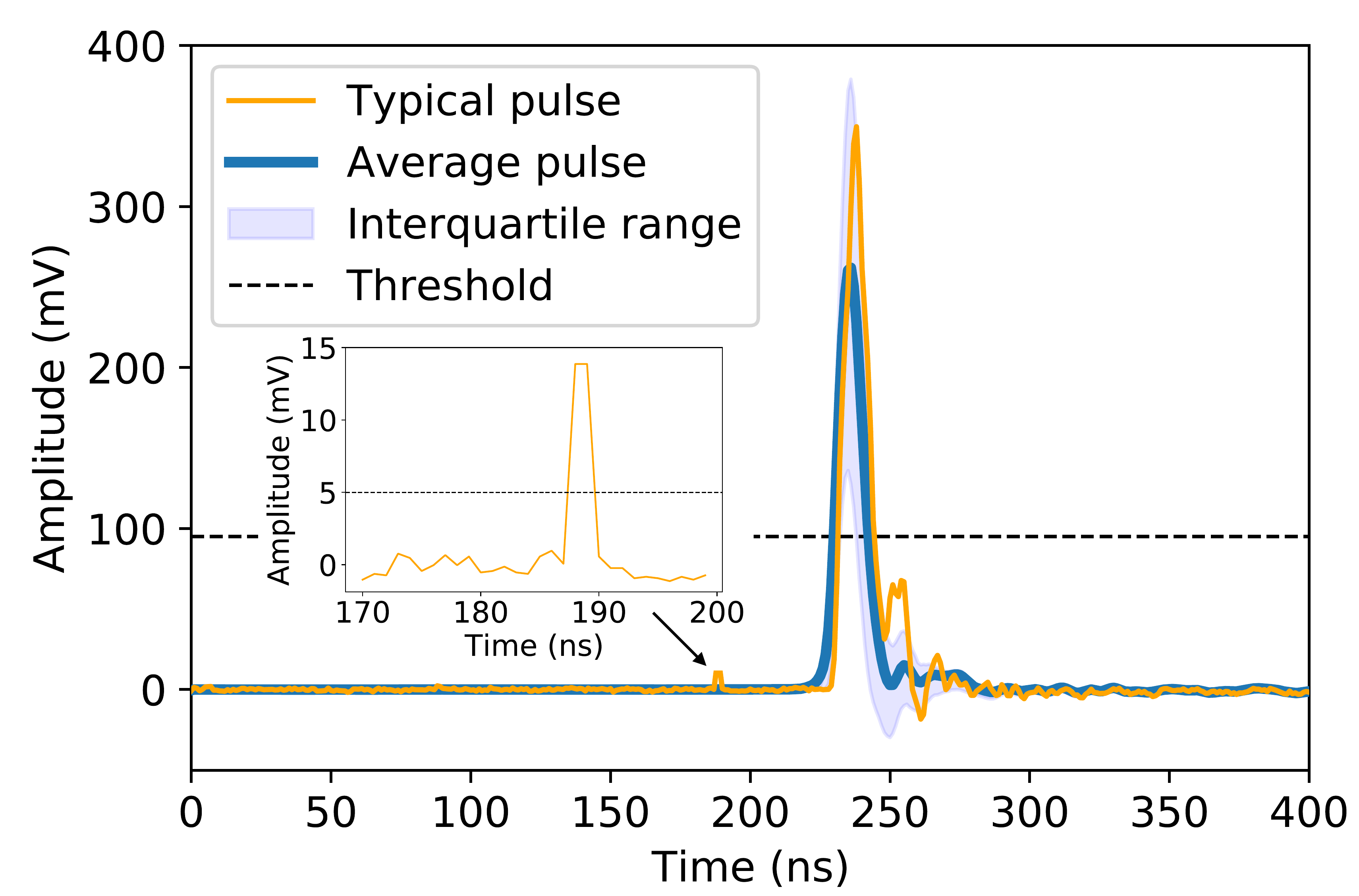}
    \caption{Typical~(orange line) and averaged~(blue line) pulses of PMT-1.
    The light blue region shows the interquartile range. The horizontal dashed line shows the threshold of data acquisition.
    The inset shows an example of a pulse of single photoelectron.
    }
    \label{fig:wf}
  \end{center}
\end{figure}

\begin{figure}[t]
  \begin{center}
    \includegraphics[width=12.0cm]{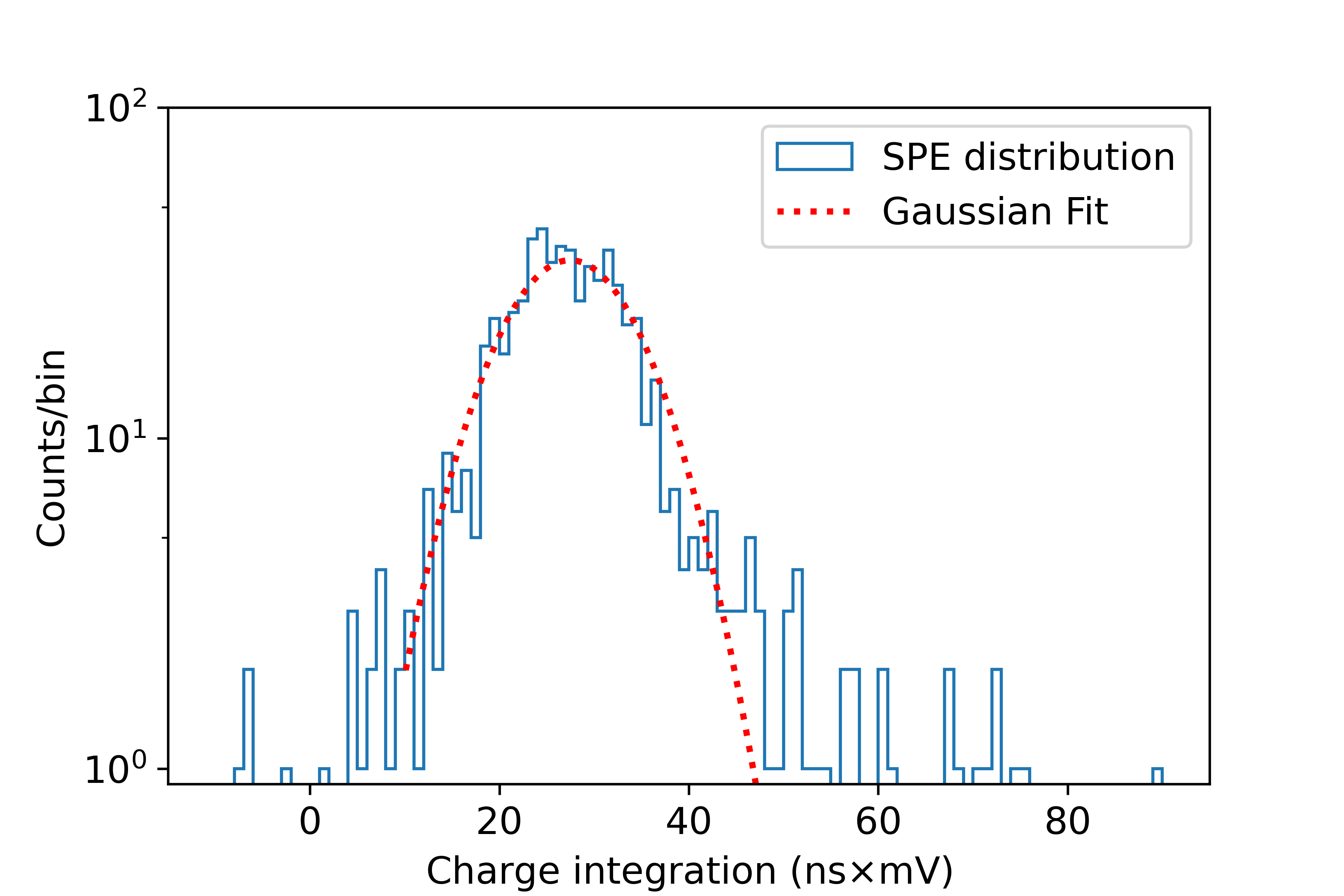}
    \caption{Typical charge distribution of SPE pulses.
    The distribution is fitted with a Gaussian function~(red dotted curve).
}
    \label{fig:pe}
  \end{center}
\end{figure}

\begin{figure}[t]
  \begin{center}
    \includegraphics[width=\linewidth]{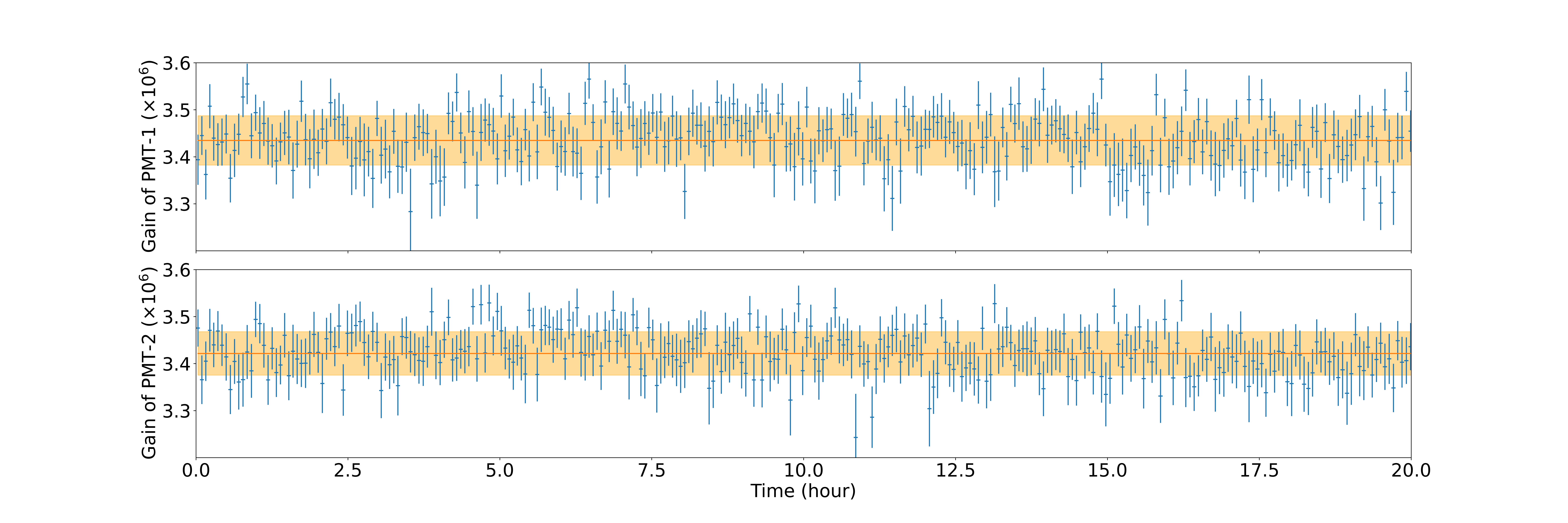}
    \caption{Gains of PMT-1 and PMT-2 during the measurement. The blue points show the gain of PMTs with the statistical uncertainty. The orange line and band show the mean and the range of its standard deviation. The gains are stable within $\pm1.5\%$ and this does not affect to the observed number of photoelectrons. }
    \label{fig:gain}
  \end{center}
\end{figure}

The stability of the PMT gains was monitored throughout the measurement by observing single photoelectron~(SPE) dark signals.
Figure \ref{fig:wf} shows a typical pulse recorded in Period A.
A fast ($\sim$20~ns of full width) pulse of scintillation light which represents the $\mathrm{CF_{4}}$'s short decay time is observed.
This figure shows an average pulse after excluding saturating pulses which were $\alpha$~particles from the source.
In this experiment, the rate of the cosmic muon is 1/100 of the signal, and the typical deposit energy is as small as 20~keV, which is negligible.
SPE pulses from accidental dark signals were searched
for before the triggered pulse timing with a software threshold of $+5$~mV. 
A typical SPE pulse is shown in the inset of Figure~\ref{fig:wf}. 
The charge for each pulse was calculated by integrating the pulse for $\pm 10$~ns around the pulse peak.
Figure~\ref{fig:pe} shows the charge distribution of SPE pulses.
The peaks are fitted with Gaussian functions and the mean values are used to calculate 
the charges of the SPE pulses, taking account of the termination resistor~(50~$\Omega$). 
Calculated SPE charges are defined as $Q_\mathrm{SPE-1}$ and $Q_\mathrm{SPE-2}$~[C] for PMT-1 and PMT-2, respectively. 
Figure~\ref{fig:gain} shows the time dependence of gains calculated as $Q_\mathrm{SPE-1,2}/e$, where $e$ is the elementary charge.
It demonstrates the stability of the PMT gains throughout the measurement. 
The fluctuations in PMT gains were found to be less than $\pm1.5\%$. 
The non-linearity of PMTs and electronics were evaluated individually with LED calibration. 
It was confirmed that the dynamic range was much larger than the light increase. 
The effect of non-linearity is negligible to the considered gain fluctuations.

\section{Results} \label{s3_results}

\subsection{Temperature dependence of light output} \label{sec:res:temp}
\begin{figure}[t]
  \begin{center}
    \includegraphics[width=10.0cm]{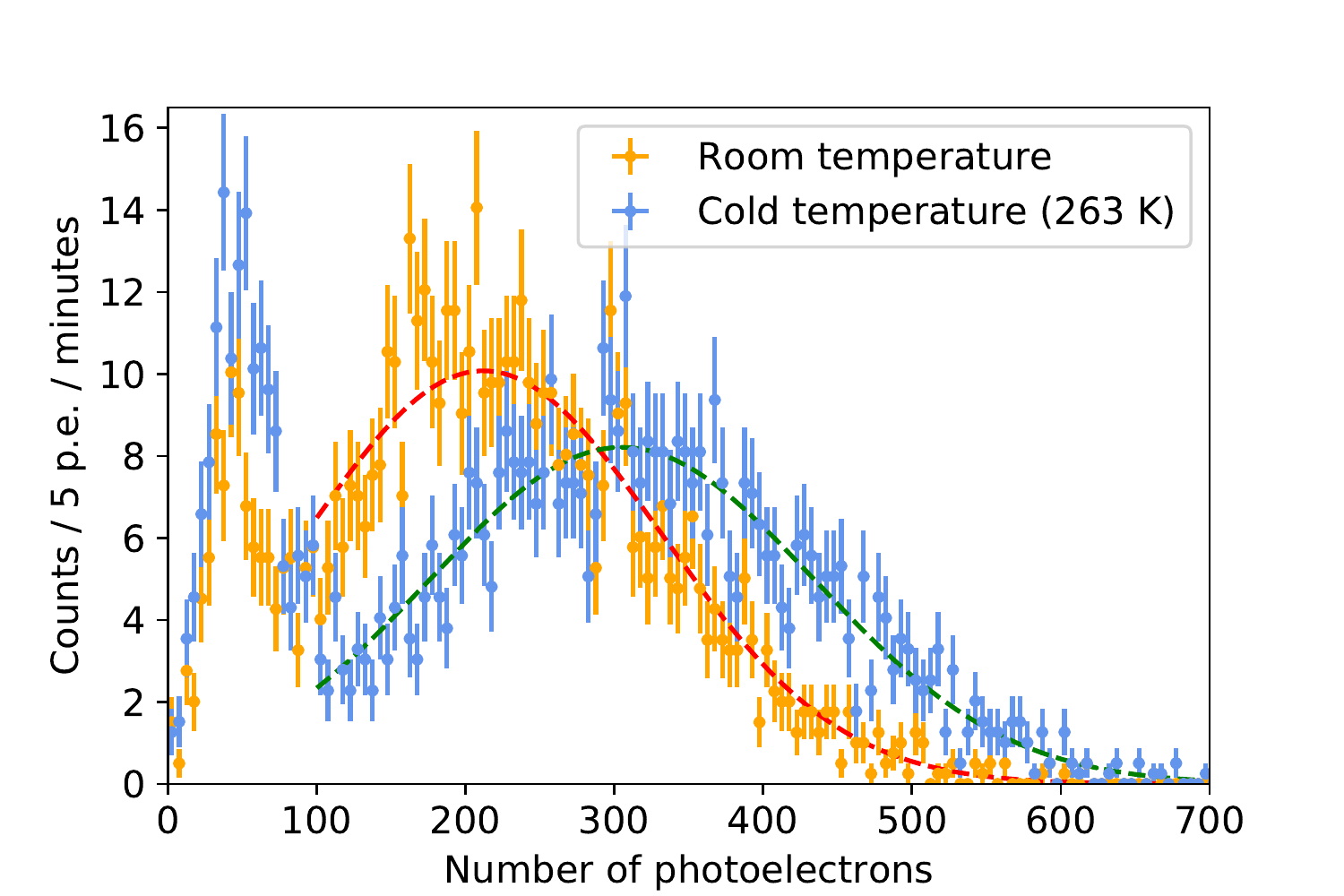}
    \caption{Typical spectra for $\gamma$-rays~($59.5$~keV). One at the room temperature is drawn with orange histogram and the other at the cool temperature~($263$~K) is drawn with blue histogram. 
    They are fitted with Gaussian function~(dotted lines). 
    In these spectra, the means of the Gaussian function are $211.9$~P.E. and $305.0$~P.E., respectively.
    The increase of emitted number of photoelectrons is clearly observed.}
    \label{fig:hist}
  \end{center}
\end{figure}

The number of photoelectrons for the observed 
event
is calculated from the waveforms of two PMTs. 
The baseline calculated from the initial $200$~ns of the waveform were subtracted from the waveform and the charge was calculated by integrating the waveform taking account of the termination resistor.
Obtained charges from PMT-1 and PMT-2 are referred to as $Q_\mathrm{PMT-1}$~[C] and $Q_\mathrm{PMT-2}$~[C], respectively. 
The total numbers of observed photoelectrons were obtained as, $N_\mathrm{total} = Q_\mathrm{PMT-1}/Q_\mathrm{SPE-1} +Q_\mathrm{PMT-2}/Q_\mathrm{SPE-2}$.
Figure \ref{fig:hist} shows a typical spectrum of  $N_\mathrm{total}$  by the 59.5~keV $\gamma$-rays in stable Periods (A and C). 
The shift of the peak position  at different temperatures is clearly observed. 
It demonstrates the high light yield at the cold temperature. 
The peaks are fitted with Gaussian function 
and the means of the functions for Periods A and C functions are 211.9~P.E. and 305.0~P.E., respectively.
   
Top plot of Figure~\ref{fig:pesum} clearly shows the increase of light yield of CF$_4$ at cold temperatures. 
\begin{figure}[t]
  \begin{center}
    \includegraphics[width=0.8\hsize]{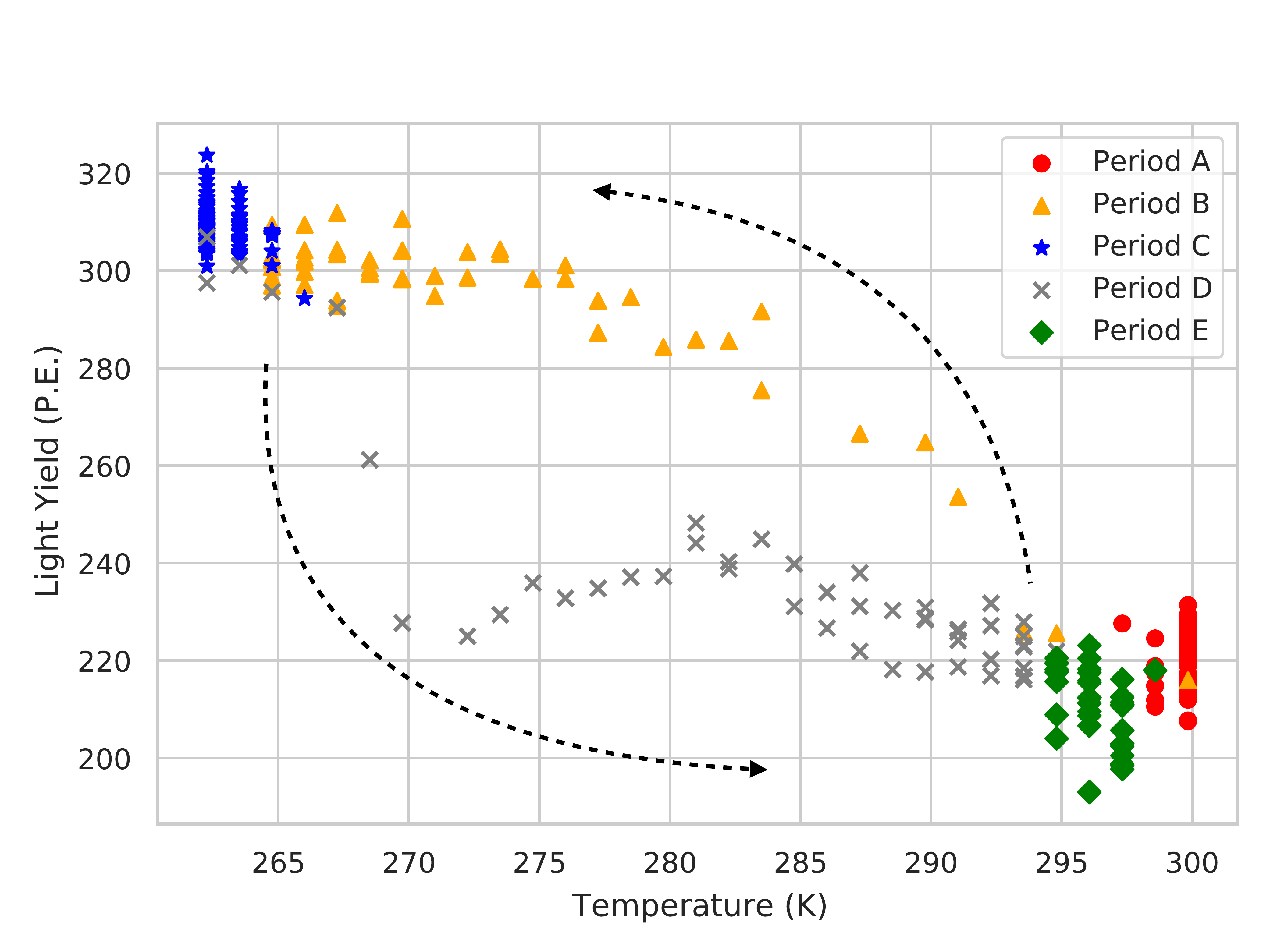}
    \caption{Measured correlation between the gas temperature and the light yield. The red circles, orange triangles, blue stars, gray crosses, and green squares corresponds the data taken in Periods A, B, C, D, and E, respectively. The arrow illustrates the time order of the measurement.
    }
    \label{fig:tvsl}
  \end{center}
\end{figure}
Table~\ref{tab:res} summarizes the light yields, resolutions for $59.5$~keV $\gamma$-ray, and gas temperatures during Periods A, C, and E. 
\begin{table}[t]
\centering
\caption{Summary of the light yield, the energy resolution, and the temperature measured in stable periods.
The resolutions are calculated for $59.5$~keV $\gamma$-ray. Uncertainties shown in this table are only statistical uncertainty.}
\label{tab:res}
\begin{tabular}{|ccccc|}
\hline
Period & Light yield~[P.E.]& Relative ratio to Period A & Resolution [\%($\sigma$)] & Gas temperature~[K] \\ \hline
A      & $219.5 \pm  5.8$  & $1$ & 51.1   & $299.7 \pm 0.5$           \\
C      & $309.1 \pm  4.5$  & $1.41 \pm 0.04$ & 41.6 & $262.8 \pm 0.8$          \\
E      & $210.9 \pm  7.6$  & $0.96 \pm 0.04$ & 52.4 & $296.3 \pm 1.0$          \\
\hline
\end{tabular}
\end{table}
In Period A, the light yield for $59.5$~keV $\gamma$-ray is measured to be $219.5\pm5.8~\mathrm{P.E.}$ with an energy resolution of $51.1\%$~($\sigma$). 
The uncertainty is calculated by quadratically summing the statistical uncertainty and the systematic uncertainty described in Section~\ref{sec_sys}.
In Period~C, the observed light yield increased to $309.1\pm4.5~\mathrm{P.E.}$
The ratio and difference between these two results are $1.41\pm0.04$ and $89.6\pm7.3~\mathrm{P.E.}$, respectively. 
The energy resolution improved from $51.1\%$ to $41.6\%$. 
In Period E, the light yield was measured to be $210.9\pm7.6~\mathrm{P.E.}$ 
The ratio and difference between Period A and E are~$0.96\pm0.04$ and~$-8.6\pm9.6~\mathrm{P.E.}$, respectively. The light yield in Period E is consistent with the one in Period A within uncertainties. 
Hence, this reproducibility of light yield demonstrates that the experimental data were properly obtained.

Figure~\ref{fig:tvsl} shows light yield as a function of the detector temperature. 
A hysteresis was observed between   
Periods B and D.
This behavior indicates a delay in the thermal conduction
of the gas from the copper plate.
There is another possibility that the light yield may have decreased due to a small amount of out-gas around the melting point even after water removal by molecular sieves.
For these reasons, we focus on the stable Periods, A, C, and E, for the discussion of the light yield.

\subsection{Systematic uncertainties} \label{sec_sys}
The observed number of photoelectrons $N_\mathrm{obs.}$ is represented as
\begin{equation}
N_\mathrm{obs.} = N_\mathrm{orig.} \times \Omega_\mathrm{eff.} \times C_\mathrm{gain} \times \varepsilon_\mathrm{QE},
\end{equation}
where $N_\mathrm{orig.}$ is the number of originally generated photons, $\Omega_\mathrm{eff.}$ is the effective solid angle of the PMTs considering reflections, $C_\mathrm{gain}$ is the PMT gain, and $\varepsilon_\mathrm{QE}$ is the quantum efficiency of the PMT.
Table~\ref{tb:sys} summarizes the systematic uncertainties for the $N_\mathrm{Total}$. 
\begin{table}[t]
    \begin{center}
    \caption{The summary of systematic uncertainties for $N_\mathrm{orig.}$. The details are described in the main text.}
        \label{tb:sys}
        \begin{tabular}{|cc|}
            \hline Item & Systematic uncertainty~[$\%$] \\ 
            \hline
            PMT gain & $\pm 1.5 $ \\
            PMT Q.E. & $\pm 5.0 $ \\
            Reproducibility & $\pm 4.0$ \\
            \hline
            Total & $\pm 6.6$ \\
            \hline
        \end{tabular}
    \end{center}
\end{table}
In this study, the $\Omega_\mathrm{eff.}$ is invariant during the measurement while the other three parameters affect $N_\mathrm{obs}$.
The uncertainty of the PMT gain, $C_\mathrm{gain}$, is evaluated as $\pm1.5\%$ of the standard deviation of the distribution of the PMT gain shown in Figure~\ref{fig:gain}. 
The uncertainty of $\varepsilon_\mathrm{QE}$ was reported in the previous study to be $\pm5\%$ at maximum~\cite{pmt_qe}.
The reproducibility is estimated at $\pm4.0\%$ from the difference between the light yields in two room temperature Periods, A and E.

These uncertainties can be judged to be independent from each other, thus the total systematic uncertainty is evaluated to be $\pm 6.6\%$.
This systematic uncertainty and measured statistical uncertainty are both small relative to the observed increases in $N_\mathrm{obs.}$, and we conclude that a significant increase in $N_\mathrm{orig.}$, or the light yield, has been observed at cold temperatures.
\section{Conclusion and future prospects}\label{s5_conclusion}
A temperature dependence of a gaseous scintillator, CF$_4$, was studied at $300$~K and $263$~K. 
The light yield of CF$_4$ was found to increase
by $(41.0\pm4.0_{\rm stat.}\pm6.6_{\rm syst.})\%$ and the energy resolution enhanced when the $\mathrm{CF}_4$ gas is cooled to $263$~K. 
As for the future prospects, this high light yield at cold temperatures enables detectors using $\mathrm{CF_{4}}$ gas to improve their energy resolution and to lower their threshold.

\section*{Acknowledgment} \label{acknowledgement}
This work was supported by KAKENHI Grant-in-Aids~(18K18768, 21K18628, 26104005, 19H05806, 19J20418, 18H05536, and 21K13942). This work is also partially supported by the joint research program of the Institute for Cosmic Ray Research~(ICRR), the University of Tokyo.

\bibliographystyle{JHEP}
\bibliography{main.bib}

\end{document}